\documentclass[print]{ieeecolor}
\usepackage{generic}
\usepackage{cite}
\usepackage{amsmath,amssymb,amsfonts}
\usepackage{algorithmic}
\usepackage{graphicx}
\usepackage{textcomp}
\usepackage{url}
\usepackage{booktabs}
\usepackage{subfig}
\usepackage{amsmath}
\usepackage{svg}
\usepackage{booktabs}
\usepackage{comment}
\usepackage{nidanfloat}
\usepackage{balance}
\usepackage{hyperref}
\usepackage{array}
\usepackage{multirow}
\usepackage{xspace}
\usepackage{amsfonts}

\usepackage{tikz}
\usepackage{bm}
\usetikzlibrary{shapes,arrows,fit,positioning,shapes.geometric}

\begin{document}
\title{Universal Physiological Representation Learning with Soft-Disentangled Rateless Autoencoders}

\author{
Mo Han, 
Ozan {\"O}zdenizci,
Toshiaki Koike-Akino,
Ye Wang,
and Deniz Erdo{\u{g}}mu{\c{s}}
\thanks{M.~Han, O.~\"{O}zdenizci, and D.~Erdo\u{g}mu\c{s} are with Cognitive Systems Laboratory, Department of Electrical and Computer Engineering, Northeastern University, Boston, MA 02115, USA. E-mail: \{han, oozdenizci, erdogmus\}@ece.neu.edu.}%
\thanks{T.~Koike-Akino and Y.~Wang are with Mitsubishi Electric Research Laboratories (MERL), Cambridge, MA 02139, USA. E-mail: \{koike, yewang\}@merl.com.}%
\thanks{M.~Han was an intern at MERL during this work. O.~{\"O}zdenizci and D.~Erdo{\u{g}}mu{\c{s}} are partially supported by NSF (IIS-1149570, CNS-1544895, IIS-1715858), DHHS (90RE5017-02-01), and NIH (R01DC009834).}%
}

\maketitle

\begin{abstract}
Human computer interaction (HCI) involves a multidisciplinary fusion of technologies, through which the control of external devices could be achieved by monitoring physiological status of users.
However, physiological biosignals often vary across users and recording sessions due to unstable physical/mental conditions and task-irrelevant activities.
To deal with this challenge, we propose a method of adversarial feature encoding with the concept of a Rateless Autoencoder (RAE), in order to exploit disentangled, nuisance-robust, and universal representations.
We achieve a good trade-off between user-specific and task-relevant features by making use of the stochastic disentanglement of the latent representations by adopting additional adversarial networks.
The proposed model is applicable to a wider range of unknown users and tasks as well as different classifiers.
Results on cross-subject transfer evaluations show the advantages of the proposed framework, with up to an $11.6\%$ improvement in the average subject-transfer classification accuracy.
\end{abstract}


\section{Introduction}

\IEEEPARstart{H}{uman} computer interaction (HCI)~\cite{Jerritta2011HCI} is a fundamental technology enabling machines to monitor physiological disorders, to comprehend human emotions, and to execute proper actions, so that users can control external devices through their physiological status in a safe and reliable fashion. 
To measure traditional physiological biosignals such as electrocardiogram (ECG)~\cite{ECG}, electromyography (EMG)~\cite{han2020HANDS} and electroencephalography (EEG)~\cite{petrantonakis2009emotion}, either implanted or surface electrodes and their frequent calibration are necessary, reducing user comfort while increasing the overall expense. 
Recently, novel wearable sensors such as wrist-worn devices were developed for accurately measuring physiological signals~\cite{Birjandtalab2016NonEEG1,Amiri2016NonEEG2,Cogan2014NonEEG3,giakoumis2013subject,giannakakis2019review,ozdenizci2018time} (e.g., arterial oxygen level, heart rate, skin temperature, etc.) in comfortable and effective manners. 
Utilizing these non-EEG physiological biosignals can effectively increase the system convenience during data collection with less expense.

One major challenge of physiological status assessment lies in the problem of transfer learning caused by the variability in biosignals across users or recording sessions due to the unstable mental/physical conditions and task-irrelevant disturbances. 
Addressing biosignal datasets collected from a narrow amount of subjects, transfer learning methods~\cite{Fazli2009transfer4,Morioka2015transfer1,Yin:2019,Chen:2019} are applied to build strong feature learning machines to extract robust and invariant features across various tasks and/or unknown subjects.
Particularly, adversarial transfer learning~\cite{ozdenizci2020learning, ozan2019adversary, ozdenizci2019advesarial, han2020adversarial, han2020DA-cAE, Edwards2015adversarial, wu2020modality, sun2020adversarial, Fader2017adversarial} demonstrated impressive results in constructing such discriminative feature extractors.
Traditional adversarial transfer learning works aim to extract latent representations universally shared by a group of attributes using adversarial inference, where a discriminative network is trained adversarially towards the feature extractor in order to differentiate universal features from various attributes. 
However, in most existing approaches, the adversarial training scheme is usually applied indiscriminately on the whole feature group when extracting cross-attribute latent representations, which inevitably leads to the loss of attribute-discriminative information.
Therefore, rather than using only one adversarial discriminator to merely preserve shared cross-attribute features, we train two additional adversarial discriminators jointly with the feature extractor, so that the physiological features could be disentangled into two counterparts representative of subject and task associated information respectively.
In this way, the variability in both subject and task space can be better accounted for.

As a commonly used feature extractor framework for transfer learning, autoencoders (AE)~\cite{Tschannen2018autoencoder, wen2018featureEncoder, yang2019conditionalAE} can learn latent representations with a dimensionality typically much smaller than the input data, which is known as a ``bottleneck'' architecture, while capturing key data features to enable data reconstruction from the latent representation.
A challenging problem in dimensionality reduction is to determine an optimal feature dimensionality which sufficiently captures latent information that is essential for particular tasks. 
To address this issue, the Rateless Autoencoder (RAE)~\cite{Koike2020RLAE} was proposed to enable the AE to seamlessly adjust feature dimensionality through its rateless property, while not requiring a fixed structure of bottleneck.
To realize such flexibility in the latent space, RAE implements a probabilistic latent dimensionality which is stochastically decreased through dropout during training, where a non-uniform dropout rate distribution is imposed to the bottleneck structure.

In this work, we propose a method of adversarial feature extractor in order to exploit soft-disentangled universal representations, extended from~\cite{han2020adversarial} and \cite{han2020DA-cAE}, where the concept of RAE is newly introduced.
Unlike traditional feature learning frameworks ignoring the specificity of either task calibrations or target subjects, the proposed model is applicable to a wider range of unknown individuals and tasks. 
Our contributions are summarized as follows:
\begin{itemize}
	\item We complementarily use two additional adversarial networks, i.e., adversary and nuisance blocks, to disentangle and re-organize the latent representations.
	\item The rateless trade-off between subject-specific and task-relevant features is exploited by stochastically attaching adversary and nuisance blocks to the encoder.
	\item Different dropout strategies of the disentangled adversarial RAE are discussed.
	\item Empirical assessments were performed on a publicly available dataset of physiological biosignals for measuring human stress level through cross-subject evaluations with various classifiers.
	\item Comparative experiments on multiple model setups including traditional autoencoder and adversarial methods are evaluated.
	\item We demonstrate the remarkable advantage of the proposed framework, achieving up to an $11.6\%$ improvement in subject-transfer classification accuracy.
\end{itemize}

\section{Methodology}

\begin{figure*}
\centering
\subfloat[Conditional autoencoder (cAE)]{\includegraphics[width=0.49\textwidth]{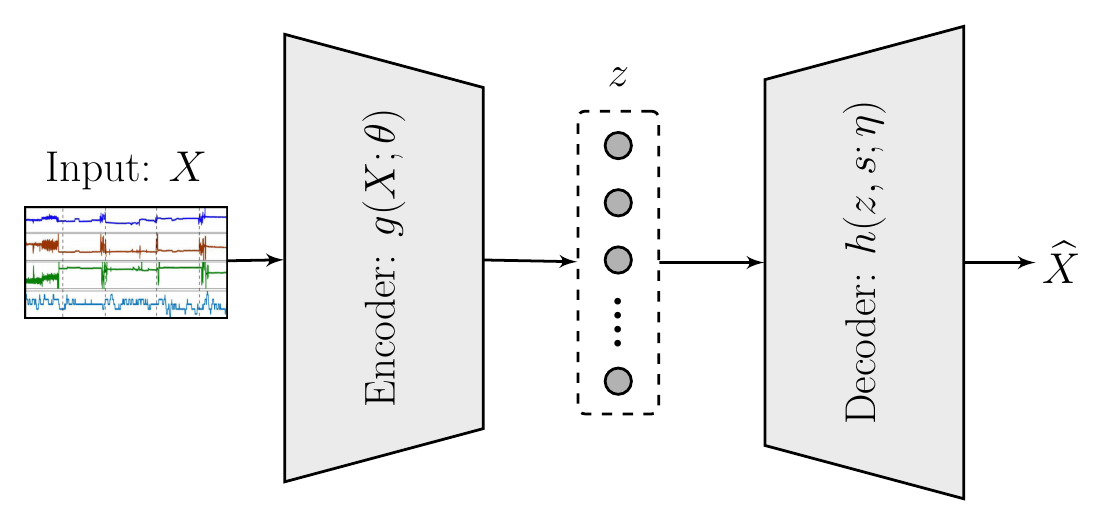}}
\hfill
\subfloat[Conditional rateless autoencoder (cRAE)]{\includegraphics[width=0.49\textwidth]{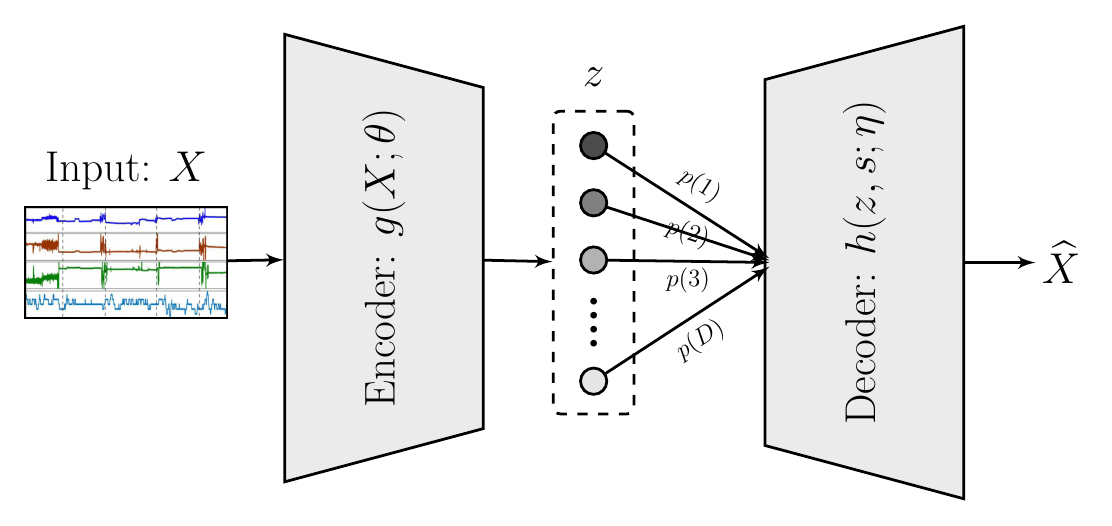}}
\caption{(a) Conditional autoencoder (cAE): an encoder-decoder pair where the encoder estimates latent $z=g(X;\theta)$ with parameters $\theta$, and the decoder estimates reconstructed input signals $\hat{X}=h(z,s;\eta)$ with parameters $\eta$, using the latent $z$ and conditioning variable $s$.
When decoder is $h(z;\eta)$, it reduces to a traditional autoencoder (AE).
(b) Conditional rateless autoencoder (cRAE): a probabilistic cAE model with a stochastic bottleneck where $d$th latent representation node is assigned with dropout probability rates $p(d)$, such that the conditional decoder takes a subset of the latent units as input.}
\label{fig:AEandRAE}
\end{figure*}

\begin{figure*}
\centering
\subfloat[Disentangled adversarial autoencoder (DA-cAE)]{\includegraphics[width=0.47\textwidth]{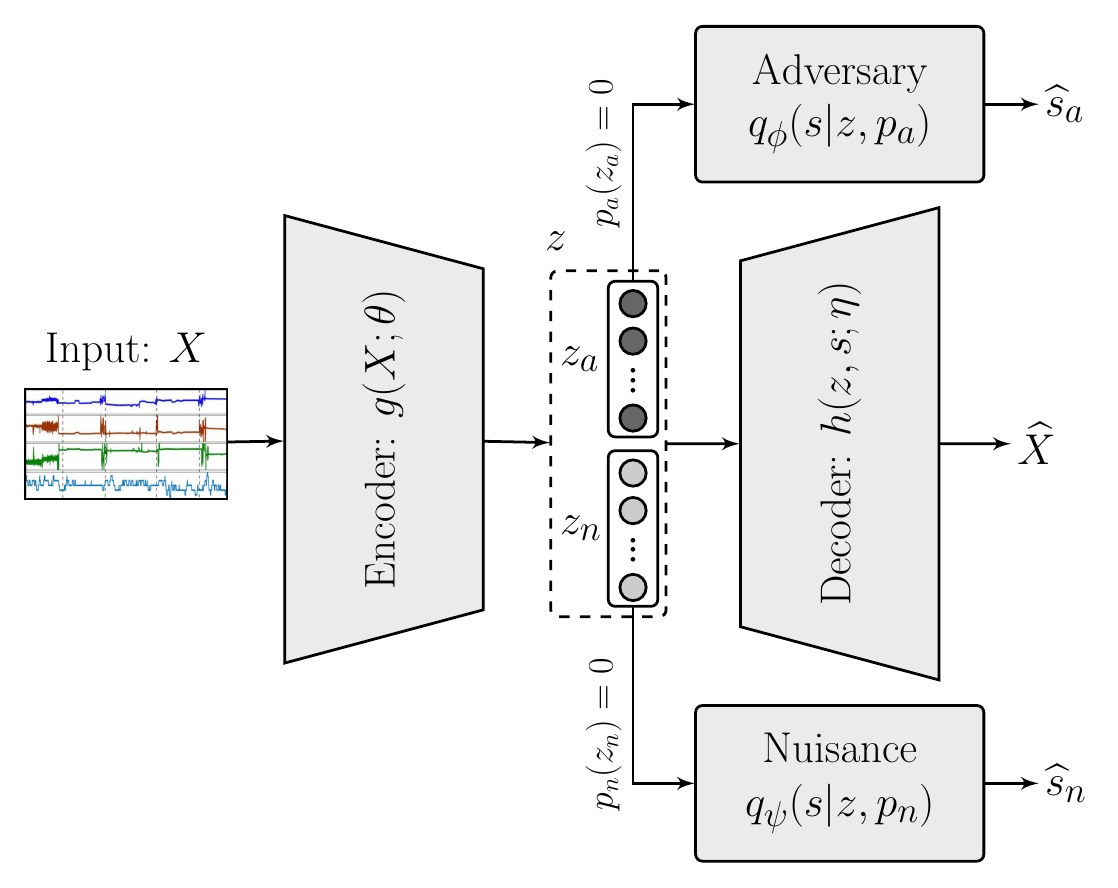}}
\hfill
\subfloat[Disentangled adversarial rateless autoencoder (DA-cRAE)]{\includegraphics[width=0.52\textwidth]{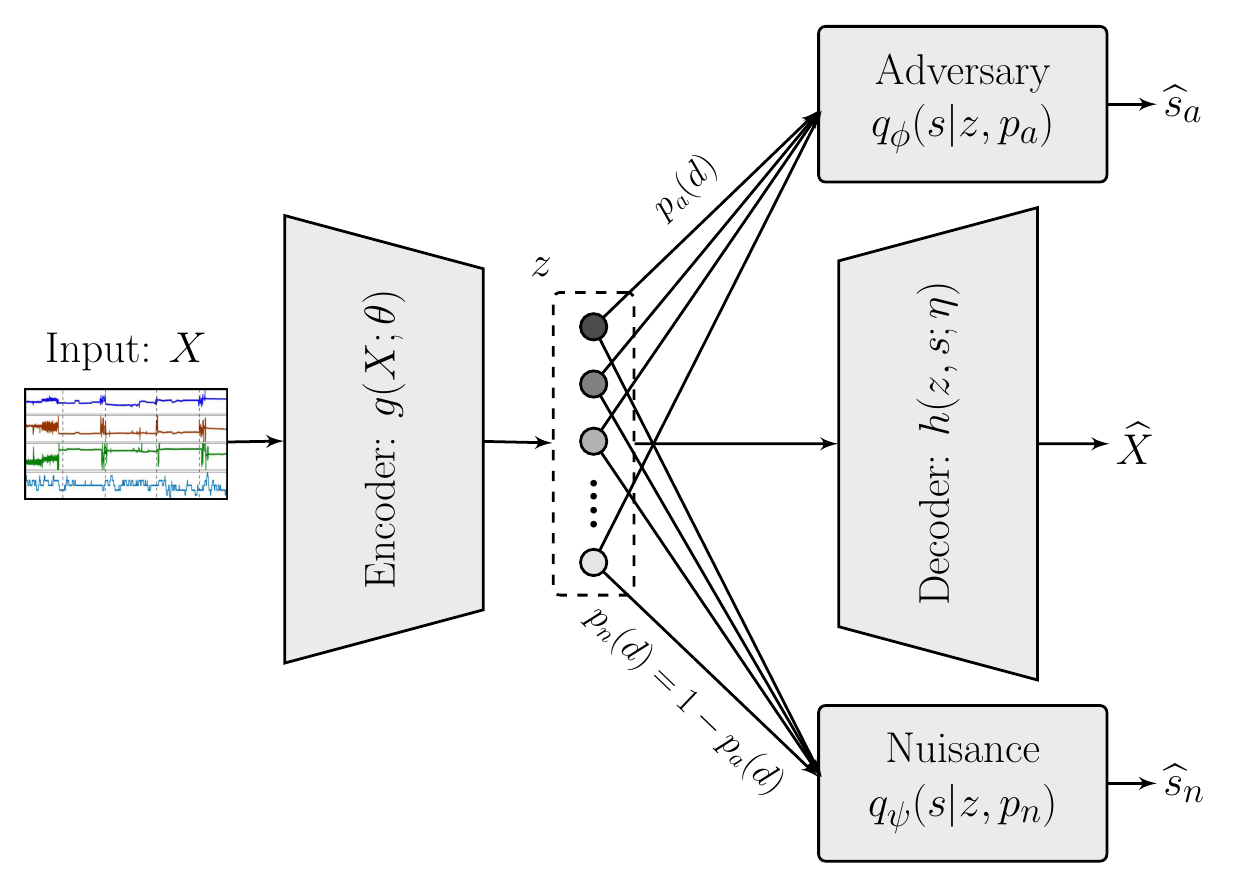}}
\caption{Disentanlgled adversarial autoencoder for nuisance-robust transfer learning.
(a) Disentangled adversarial conditional autoencoder (DA-cAE) with hard split: a deterministic disentangled universal latent representation learning model where $z$ is partitioned into sub-parts $z_a$ and $z_n$ which are adversarially trained to be $s$-invariant (i.e., $z_a$ used as an input to an adversary network) and $s$-variant (i.e., $z_n$ used as an input to a nuisance network) respectively.
(b) Disentangled conditional rateless autoencoder (DA-cRAE) with soft split: a cRAE model with soft disentanglement, where the adversary and nuisance network inputs are determined through the stochastic bottleneck architecture with probabilities $p_a(d)$ and $p_n(d)=1-p_a(d)$ respectively for the $d$th latent node. 
}
\label{fig:DRAE}
\end{figure*}

\subsection{Notation and Problem Description}

We define $\{(X_i,y_i,s_i)\}_{i=1}^{n}$ as a labeled data set, where $X_i \in \mathbb{R}^{C}$ is the input data vector recorded from $C$ channels of trial $i$, $y_i \in {\{0,1,\ldots,L-1\}}$ is the  class label of user task/status among $L$ classes, and $s_i \in {\{1,2,\ldots,S\}} $ is the user identification (ID) index among $S$ subjects.
The task/status $y$ is assumed to be marginally independent with respect to subject ID $s$, and the physiological signal is generated dependently on $y$ and $s$, i.e., $X\sim p(X|y,s)$. 
The aim is to construct a model to estimate the task/status label $y$ given an observation $X$, where the model is generalized across the variability of subject $s$, which is considered as a nuisance variable associated with transferring the feature extraction model.

\subsection{Rateless Autoencoder (RAE)}

AE is a well-known feature learning machine which includes a network pair of encoder and decoder, as shown in Fig.~\ref{fig:AEandRAE}(a). 
The encoder packs data features into a latent representation $z$, while the decoder intends to re-construct the input data $X$ based on the latent representation $z$. 
AE structures are typically bottleneck architectures, where the dimensionality $D$ of representation $z$ is lower than the dimensionality of input data $X$, and the latent variables should contain adequate features capable of reconstructing the original data through its corresponding decoder network. 
A challenging problem in such a dimensionality reduction is to decide an optimal feature dimensionality which captures sufficient latent representations that are essential for specific tasks.

RAE~\cite{Koike2020RLAE} is an AE family providing a rateless property that enables the AE to seamlessly adjust feature dimensionality. 
Unlike a conventional AE with a deterministic bottleneck architecture, the RAE employs a probabilistic bottleneck feature $z$ whose dimensionality $D$ is stochastically reduced through dropout. 
Particularly, RAE imposes a specific dropout rate distribution that varies across the $D$ nodes of representation $z$. 
For example, as depicted in Fig.~\ref{fig:AEandRAE}(b), the RAE encoder generates latent variables $z$ of dimension $D$ which are randomly dropped out at a probability of $p(d)$ for node $d \in {\{1,2,\ldots,D\}} $, resulting in an effective latent dimensionality of $\bar{D} = \sum_{d=1}^{D}{(1-p(d))}$.
RAE is regarded as an ensemble method which jointly exploits all different AEs having a latent dimension of $d$ from $1$ to $D$.
It is hence more insensitive to the choice of the dimensionality parameter.

In our method, we make use of the RAE concept to realize a good trade-off between task-related features and person-discriminative information by attaching new adversary and nuisance blocks to the representation $z$ through different dropout strategies, with $z$ fed into the decoder without dropout.
A soft-disentangled feature extractor is first trained based on the rateless conception, and a task classifier is then learned for the final discriminative model utilizing the features extracted from the pre-trained (frozen) feature encoder.

\subsection{Disentangled Adversarial Transfer Learning with RAE}

\subsubsection{Disentangled Feature Extractor}

In~\cite{han2020adversarial} and \cite{han2020DA-cAE}, disentangled feature extraction method was proposed to improve subject-transfer performance.
As shown in Fig.~\ref{fig:DRAE}(a), the features $z$ are divided into two parts of $z_a$ and $z_n$, which are intended to conceal subject-invariant and subject-specific information, respectively.
Despite the gain of the disentangled method, determining the split sizes of $z_a$ and $z_n$ is still challenging.
In this paper, we extend the method with soft disentanglement motivated by RAE as shown in Fig.~\ref{fig:DRAE}(b), to mitigate the sensitivity of the splitting parameter.

For implementing the soft-disentangled adversarial transfer learning, encoder output $z$ is forwarded into two additional units, the \textit{adversary network} and \textit{nuisance network}, with different dropout rate distributions.
As illustrated in Fig.~\ref{fig:DRAE}(b), the dropout rate distributions of representation $z$ to the adversary network and nuisance network are designed as $p_a(d)$ and $p_n(d)=1-p_a(d)$, respectively. 
Complete latent representation $z$ is further fed into the decoder $h(z,s;\eta)$ without any dropout.  
Through the stochastic disentangling, the representations $z$ are re-organized into two sub-parts related to task and subject respectively: upper feature units with lower $p_a(d)$ (higher $p_n(d)$) to adversary network aim to conceal more subject information regarding $s$, while lower units with lower $p_n(d)$ (higher $p_a(d)$) to nuisance network are designed to include more subject-related features. 
By dissociating the nuisance variable from task-related feature in a more clear way, the model is extrapolated into a broader domain of subjects and tasks. 
For the input data from an unknown user, task-related features with lower $p_a(d)$ would be incorporated into the final prediction; simultaneously, the biological characteristics which are similar to known subjects could also be projected to representations with lower $p_n(d)$ as a reference.

In order to filter out the variation elements caused by $s$ from the adversary counterpart of $z$ with lower $p_a(d)$ and simultaneously maintain more task-relevant information in it, the encoder is driven to minimize the adversary likelihood of $q_\phi\left(s|z, p_a\right)$; at the same time, to embed sufficient user-discriminative features within representations with lower $p_n(d)$, the encoder is also forced to maximize the nuisance likelihood of $q_\psi\left(s|z, p_n\right)$.
The full representation $z$ from encoder is fed into the decoder with zero dropout, which is conditioned on $s$ as an additional input besides $z$, where the encoder and decoder are trained to optimize the reconstruction loss of $\hat{X}$ compared to the true input $X$. 
Therefore, the final objective function to train the proposed model structure can be written as follows:
\begin{align}
         &\mathsf{Loss}_\mathrm{RAE}(X;\eta,\theta,\psi,\phi) =
           - \mathbb{E} \big[ \log p_\eta \big(\hat{X}|g(X;\theta), s \big) \big]  \notag\\
          & \quad {} - \lambda_N \mathbb{E} \left[ \log q_\psi \left(s|z, p_n \right) \right]   
           + \lambda_A \mathbb{E} \left[ \log q_\phi \left(s|z, p_a \right) \right],
\end{align}
where the first item is the loss of decoder $\hat{X}=h(z,s;\eta)$ reconstructing inputs from $z=g(X;\theta)$, and $\lambda_A \geq 0$ and $\lambda_N \geq 0$ respectively represent the regularization weights for adversary and nuisance units in order to achieve a flexible trade-off between identification and invariance performance. 
The model will reduce to a regular conditional AE (cAE) structure when $\lambda_A = \lambda_N = 0$, which involves no stochastic bottleneck or disentangling transfer learning block. 

\subsubsection{Adversarial Training Scheme}
In addition to the training of encoder-decoder pair, at every optimization iteration, the parameters of adversary and nuisance networks are learned towards maximizing the likelihoods $q_\phi\left(s|z, p_a\right)$ and $q_\psi\left(s|z, p_n\right)$ respectively to estimate the ID $s$ among $S$ subjects. 
The parameter updates and optimizations among the encoder-decoder pair, adversary network and nuisance network are performed alternatingly by stochastic gradient descent, where the two adversarial discriminators are separately trained to minimize their corresponding cross-entropy losses.

\subsubsection{Discriminative Classifier}
An independent status/task classifier is attached to the encoder with frozen network weights pre-trained by the proposed soft-disentangled adversarial method, and then optimized utilizing the input of latent feature $z$.
The purpose of the classifier is to estimate the corresponding status/task class $y$ among $L$ categories given the physiological input $X$, where the feature $z=g(X;\theta)$ of $X$ would be first extracted ahead to the task classifier. 
Parameterized by $\gamma$, the classifier optimization is further executed by minimizing the following cross-entropy loss:
\begin{align}
\label{clsloss}
          \mathsf{Loss}_\mathrm{C}(z;\gamma) =
          \mathbb{E} \left[ -\log p_\gamma \left(\hat{y}|z \right) \right],
\end{align}
where $\hat{y}$ is the estimate of subject status/task category $y$.

\subsection{Discussion of Dropout Rate Distribution}

Within the various dropout rate distributions for the representation $z$ input to the adversary and nuisance networks (when $\lambda_A > 0$ and $\lambda_N > 0$), the stochastic bottleneck architecture includes two cases: hard split and soft split.

\subsubsection{Hard Split}
\label{Hard Split}
For the particular case when the dropout rate $p_a(d)=1-p_n(d)$ is either $0$ or $1$ for each feature node $d$, i.e., when the feature output of node $d$ is either input to the adversary network only or the nuisance network only along with decoder, the representation $z$ is hard split into two sub-parts $z_a$ and $z_n$, corresponding respectively to the adversary and nuisance blocks, as shown in Fig.~\ref{fig:DRAE}(a). 
The sub-part feature $z_a$ with $p_a(d)=0$ and $p_n(d)=1$ for $d \in z_a $ aims at preserving task-related feature information, while subject-related feature would be embedded in representation $z_n$ with $p_a(d)=1$ and $p_n(d)=0$ for $d \in z_n $. 
In this case, it reduces to a regular disentangled adversarial cAE structure (DA-cAE) with adversary and nuisance networks attached but no rateless property, as introduced in \cite{han2020adversarial, han2020DA-cAE}.

\subsubsection{Soft Split}
\label{Soft Split}
For the more generic case of soft-split representation $z$, dropout rates to adversary and nuisance blocks are arbitrary, provided that they satisfy $p_a(d) = 1- p_n(d) \in [0, 1]$ for each feature node $d \in {\{1,2,\ldots,D\}} $.
Therefore, the bottleneck architecture $z$ is soft split into adversary and nuisance counterparts stochastically according to the distribution $p_a(d)$ and $p_n(d)=1-p_a(d)$, respectively, as depicted in Fig.~\ref{fig:DRAE}(b).
This conditional RAE with soft-disentangled adversarial structure (DA-cRAE) can partly resolve the issue of hard split which requires pre-determined dimensionality for two disentangled latent vectors, whereas the proposed method can automatically consider different ratio of hard splits in a non-deterministic ensemble manner.

\subsection{Model Implementations}
Recently, neural networks and deep learning show impressive results in biosignal processing~\cite{ozdenizci2019advesarial,atzori2016deep,faust2018deep}. 
Motivated by those works, we mainly make use of neural networks to build feature extractor in the proposed model. 
However, we note that other learning frameworks without neural networks is also be able to be applied to the proposed method of soft-disentangled adversarial transfer learning.

\subsubsection{Model Architecture}
\begin{table}[t]
    \caption{Network structures, where FC($d_\mathrm{i}, d_\mathrm{o}$) is linear fully connected layer of dimensions $d_\mathrm{i}$ and $d_\mathrm{o}$ for input and output, and ReLU denotes rectified linear unit.}
    \centering
    \begin{tabular}{ c  c  } 
    \toprule 
    Encoder Network & FC($C$, $D$) $\rightarrow$ ReLU $\rightarrow$ FC($D$, $D$)\\
    Decoder Network & FC($D$, $D$) $\rightarrow$ ReLU $\rightarrow$ FC($D$, $C$) \\
    Adversary Network  & FC($D$, $S$) \\
    Nuisance Network & FC($D$, $S$) \\
    \bottomrule
    \end{tabular}
    \label{model_architectures}
\end{table}
The utilized model structure for experiment evaluations is presented in Table~\ref{model_architectures}, where representation $z$ has a dimensionality of $D$. 
The adversary and nuisance networks have a same input dimension $D$ from the latent representation and output dimension $S$ for the classification of subject IDs.
We note that we did not observe significant improvements by deepening the network or altering the number of units for our physiological biosignal dataset under test. 
To assess the robustness of the proposed soft-disentangled adversarial feature encoder, we implemented various classifiers for evluating the final task classification, including MLP, nearest neighbors, decision tree, linear discriminant analysis (LDA), and logistic regression classifiers with $L$ output dimensions for task classification.

\subsubsection{Rateless Parameters}
Representation $z$ with dimension $D=15$ is fed into adversary network and nuisance network respectively with dropout rates $p_a(d)$ and $p_n(d)$. 
For the soft-split case in Section~\ref{Soft Split}, we take $p_a(d)=( (d-1)/(D-1) )^\alpha$ and $p_n(d)=1-p_a(d)$ for $d \in {\{1,2,\ldots,D\}} $, where parameter $\alpha$ can adjust the ascent speed of dropout rate $p_a(d)$ along $d$, and we take $\alpha=3$ in the experimental assessments. 
In the implementation for hard split of Section~\ref{Hard Split}, we fix the ratio of dimensions between $z_a$ and $z_n$ to $2:1$.

\subsubsection{Comparison Model Definitions}
\label{Model Definition}
We denote AE as a baseline architecture of a regular encoder-decoder pair for feature extraction as presented in \cite{Tschannen2018autoencoder} and \cite{wen2018featureEncoder}, whose decoder is $h(z;\eta)$ without adversarial disentangling units, and cAE as a conditional AE feature extractor with decoder $h(z,s;\eta)$ conditioned on $s$ as described in \cite{yang2019conditionalAE}.
A-cAE and A-cRAE denote the cAE models with the aforementioned hard-split and soft-split bottleneck features respectively attached to the adversary network only. D-cAE and D-cRAE represent cAE with hard-split and soft-split bottleneck variables respectively linked to the nuisance network only. DA-cAE and DA-cRAE specify hard-split and soft-split representations connected to both adversary and nuisance networks respectively with decoder conditioned on $s$.
Note that the A-cAE resembles to the traditional adversarial learning methods presented in \cite{Edwards2015adversarial, wu2020modality, sun2020adversarial, Fader2017adversarial} where only one adversarial unit is adopted.

\section{Experimental Study}

\subsection{Dataset}

The proposed methodology was evaluated on a physiological biosignal dataset for assessing human stress status~\cite{Birjandtalab2016NonEEG1}, which is available online\footnote{\url{https://physionet.org/content/noneeg/1.0.0/}}.
It includes physiological biosignals of various modalities, in order to estimate $L=4$ discrete stress levels (physical stress, cognitive stress, emotional stress, and relaxation) based on data collected from $S=20$ subjects. 
The biosignals were generated from non-invasive biosensors worn on the wrist, containing heart rate, temperature, electrodermal activity, three-dimensional acceleration, and arterial oxygen level, therefore resulting in $C=7$ signal channels totally.
We further downsampled the signals to $1$~Hz in order to align all data channels. 
For each stress status, a $5$-minute long task was assigned to the subjects. 
In total, $7$ trials were executed by every subject, among which $4$ trials were the status of relaxation. 
To address the data imbalance of trials with different categories, we only utilized the first trial of relaxation status, leading to  four trials for the four stress status levels respectively and $24{,}000$ data samples in total.

\begin{figure*}[t!]
  \centering
  \includegraphics[width=\textwidth]{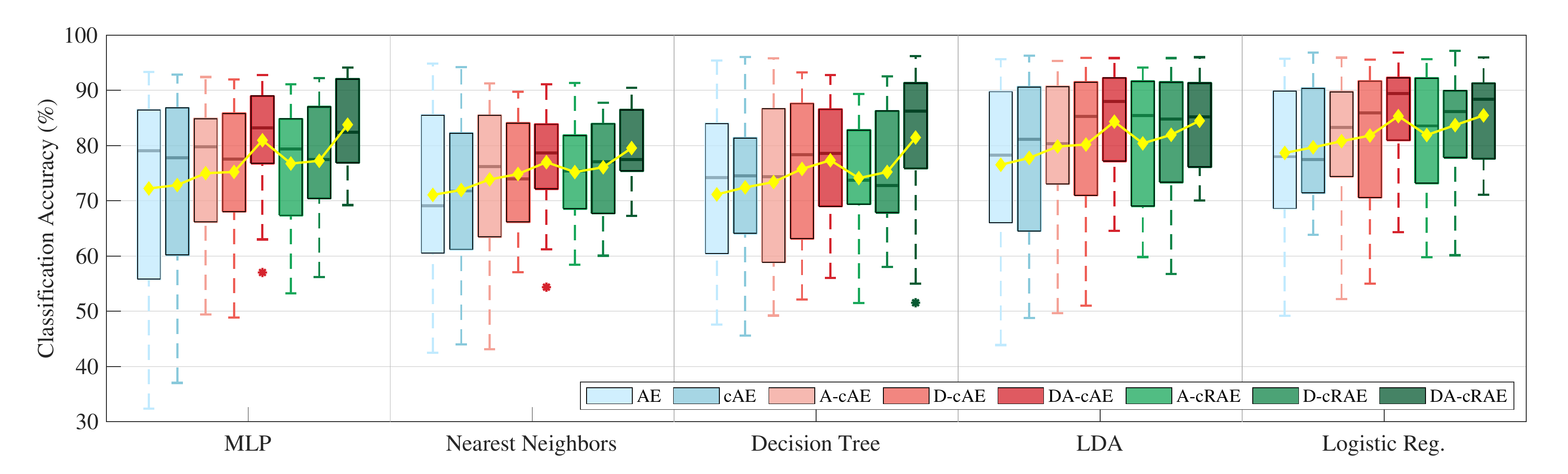}
  \caption{Transfer learning accuracies for $20$ held-out subjects of different classifiers with eight feature learning frameworks: (1) AE: baseline of regular AE with decoder $h(z;\eta)$, (2) cAE: AE with $s$-conditional decoder $h(z,s;\eta)$, (3) A-cAE: hard-split bottleneck cAE with adversary network, (4) D-cAE: hard-split bottleneck cAE with nuisance network, (5) DA-cAE: hard-split bottleneck cAE with both adversary and nuisance networks, (6) A-cRAE: soft-split bottleneck cAE with adversary network, (7) D-cRAE: soft-split bottleneck cAE with nuisance network, (8) DA-cRAE: soft-split bottleneck cAE with both adversary and nuisance networks. For each box, the central line marks the median, upper and lower bounds represent first and third quartiles, and dashed lines denote extreme values; the diamond-shape marker specifies the average.}
\label{results_box}
\end{figure*}

\subsection{Experiment Implementation}

The regularization weights $\lambda_A$ and $\lambda_N$ were chosen for the disentangled adversarial model by parameter sweep and validation.
We trained the model with different parameter combinations, and preferred the parameters producing lower accuracy of the adversary discriminator and higher accuracy of the nuisance discriminator, premised on obtaining higher cross-validation accuracy for the discriminative task classifier.

While optimizing the selection for $\lambda_A$ and $\lambda_N$, to reduce the size of parameter combinations, we first swept over $\lambda_N$ with $\lambda_A=0$; then $\lambda_N$ was fixed at its optimized value from the previous step to optimize $\lambda_A$ value. 
The adopted ranges of $\lambda_A$ and $\lambda_N$ are $\lambda_A \in \{0, 0.01, 0.05, 0.1, 0.2, 0.5 \}$ and $\lambda_N \in \{0, 0.005, 0.01, 0.05, 0.2, 0.5\}$. 
Note that the selected parameter values can be even optimized more within larger scopes by cross-validating the same model learning process.
We evaluated the model with transfer analysis of cross-subjects through a leave-one-subject-out method, where the cross-subject test data came from the left-out subject, and $90\%$ and $10\%$ of the data from the remaining subjects were randomly split as the training and validation sets respectively.

\begin{table*}[ht!]
    \caption{Optimized parameter selections with averaged cross-subject accuracies. 
    }
    \centering
    \scalebox{0.95}{
    \begin{tabular}{c c c c c c c c c c c c c c c c c c c c}
    \toprule
     & \multicolumn{3}{c}{MLP}  & \multicolumn{3}{c}{Nearest Neighbors}  & \multicolumn{3}{c}{Decision Tree} & \multicolumn{3}{c}{LDA}  & \multicolumn{3}{c}{Logistic Regression}  \\
     \cmidrule(r){2-4} \cmidrule(r){5-7} \cmidrule(r){8-10} \cmidrule(r){11-13} \cmidrule(r){14-16} \cmidrule(r){17-19}
     & $\lambda_A$ & $\lambda_N$ & avg acc & $\lambda_A$ & $\lambda_N$ & avg acc & $\lambda_A$ & $\lambda_N$ & avg acc & $\lambda_A$ & $\lambda_N$ & avg acc & $\lambda_A$ & $\lambda_N$ & avg acc \\ \toprule
    AE~\cite{Tschannen2018autoencoder,wen2018featureEncoder} & 0 & 0 & \multicolumn{1}{c|}{72.2\%} & 0 & 0 & \multicolumn{1}{c|}{71.1\%} & 0 & 0 & \multicolumn{1}{c|}{71.2\%} & 0 & 0 & \multicolumn{1}{c|}{76.5\%} & 0 & 0 & 78.7\%\\  
    cAE~\cite{yang2019conditionalAE} & 0 & 0 & \multicolumn{1}{c|}{72.9\%} & 0 & 0 & \multicolumn{1}{c|}{72.2\%} &  0 & 0 &  \multicolumn{1}{c|}{72.4\%} &  0 & 0 & \multicolumn{1}{c|}{77.8\%} & 0 & 0 & 79.7\%\\   \midrule
    A-cAE~\cite{Edwards2015adversarial, wu2020modality, sun2020adversarial, Fader2017adversarial} & 0.005 & 0 & \multicolumn{1}{c|}{75.0\%} & 0.1 & 0 & \multicolumn{1}{c|}{73.9\%} &  0.1 & 0 &  \multicolumn{1}{c|}{73.4\%} &  0.05 & 0 & \multicolumn{1}{c|}{79.8\%} & 0.05 & 0 & 80.8\%\\
    D-cAE & 0 & 0.005 & \multicolumn{1}{c|}{75.2\%} & 0 & 0.01 & \multicolumn{1}{c|}{74.9\%} & 0 & 0.01 &  \multicolumn{1}{c|}{75.8\%} & 0  & 0.2 & \multicolumn{1}{c|}{80.2\%} &  0  & 0.2 & 81.8\%\\  
    \textbf{DA-cAE}~\cite{han2020adversarial, han2020DA-cAE} & 0.01 & 0.005 & \multicolumn{1}{c|}{\bf 81.0\%} & 0.1 & 0.01 & \multicolumn{1}{c|}{\bf 77.0\%} &  0.2 & 0.01 &  \multicolumn{1}{c|}{\bf 77.3\%} &  0.2 & 0.2 & \multicolumn{1}{c|}{\bf 84.3\%} & 0.2 & 0.2 & \bf 85.3\%\\  \midrule
    A-cRAE & 0.02 & 0 & \multicolumn{1}{c|}{76.8\%} & 0.05 & 0 & \multicolumn{1}{c|}{75.2\%} &  0.05 & 0 &  \multicolumn{1}{c|}{74.1\%} &  0.1 & 0 & \multicolumn{1}{c|}{80.4\%} & 0.02 & 0 & 81.9\%\\
    D-cRAE & 0 & 0.05 & \multicolumn{1}{c|}{77.2\%} & 0 & 0.05 & \multicolumn{1}{c|}{76.1\%} & 0 & 0.1 &  \multicolumn{1}{c|}{75.2\%} & 0  & 0.05 & \multicolumn{1}{c|}{82.0\%} & 0  & 0.05 & 83.7\%\\ 
    \textbf{DA-cRAE} & 0.5 & 0.05 & \multicolumn{1}{c|}{\bf 83.8\%} & 0.5 & 0.05 & \multicolumn{1}{c|}{\bf 79.6\%} & 0.01 & 0.1 &  \multicolumn{1}{c|}{\bf 81.5\%} & 0.5 & 0.05 & \multicolumn{1}{c|}{\bf 84.5\%}  & 0.5 & 0.05 & {\bf 85.5\%}\\ 
    \bottomrule
    \end{tabular}
    }
    \label{params}
\end{table*}

\subsection{Results and Discussions}

\subsubsection{Comparative Experiments}
Accuracies of transfer analysis across $20$ held-out subjects based on different feature encoders and classifiers are presented in Fig.~\ref{results_box}, where AE, cAE, A-cAE, D-cAE, DA-cAE, A-cRAE, D-cRAE, and DA-cRAE as defined in Section~\ref{Model Definition} were trained and compared. 
Corresponding parameter settings for each case in Fig.~\ref{results_box} are displayed in Table~\ref{params}, which were selected and optimized through the aforementioned parameter optimization procedure. The model architecture is as shown in Table~\ref{model_architectures}, where feature dimension is $D=15$.

\begin{table}[t]
    \caption{Parameter optimization of MLP classifier. Accuracies for the adversary, nuisance and classifier are presented.}
    \centering
    \begin{tabular}{c c c c c c c}
    &  &  & \textbf{MLP} & \textbf{Adversary} & \textbf{Nuisance} \\ 
     & $\lambda_A$ & $\lambda_N$ & \textbf{Classifier} & \textbf{Network} & \textbf{Network} \\ \toprule
    AE & 0 & 0 & 72.2\% & 7.8\% & 5.6\% \\  \midrule
    cAE & 0 & 0 & 72.9\% & 8.5\% & 5.8\% \\  \midrule
    & 0 & 0.005 & 74.8\% & 7.7\% & 8.5\% \\
    \multirow{2}{*}{D-cRAE}  & 0 & 0.01 & 73.5\% & 12.5\% & 15.2\% \\
    & \textbf{0} & \textbf{0.05} & \textbf{77.2\%} & \textbf{10.7\%} & \textbf{19.7\%} \\
    & 0 & 0.2 & 75.6\% & 13.6\% & 16.5\% \\
    & 0 & 0.5 & 74.1\% & 12.6\% & 35.5\% \\ \midrule
    \multirow{5}{*}{\textbf{DA-cRAE}} & 0.01 & 0.05 &78.3\% & 9.4\% & 13.6\% \\
    & 0.05 & 0.05 & 77.3\% & 6.7\% & 14.6\% \\
    & 0.1 & 0.05 & 77.9\% & 5.9\% & 13.3\% \\
    & 0.2 & 0.05 & 81.5\% & 5.5\% & 12.7\% \\
    & \textbf{0.5} & \textbf{0.05} & \textbf{83.8\%} & \textbf{4.9\%} & \textbf{13.9\%} \\
    \bottomrule
    \end{tabular}
    \label{results}
\end{table}
\begin{figure}[t!]
  \centering
  \includegraphics[clip,trim=0.3cm 0 0.8cm 0.1cm, width=\linewidth]{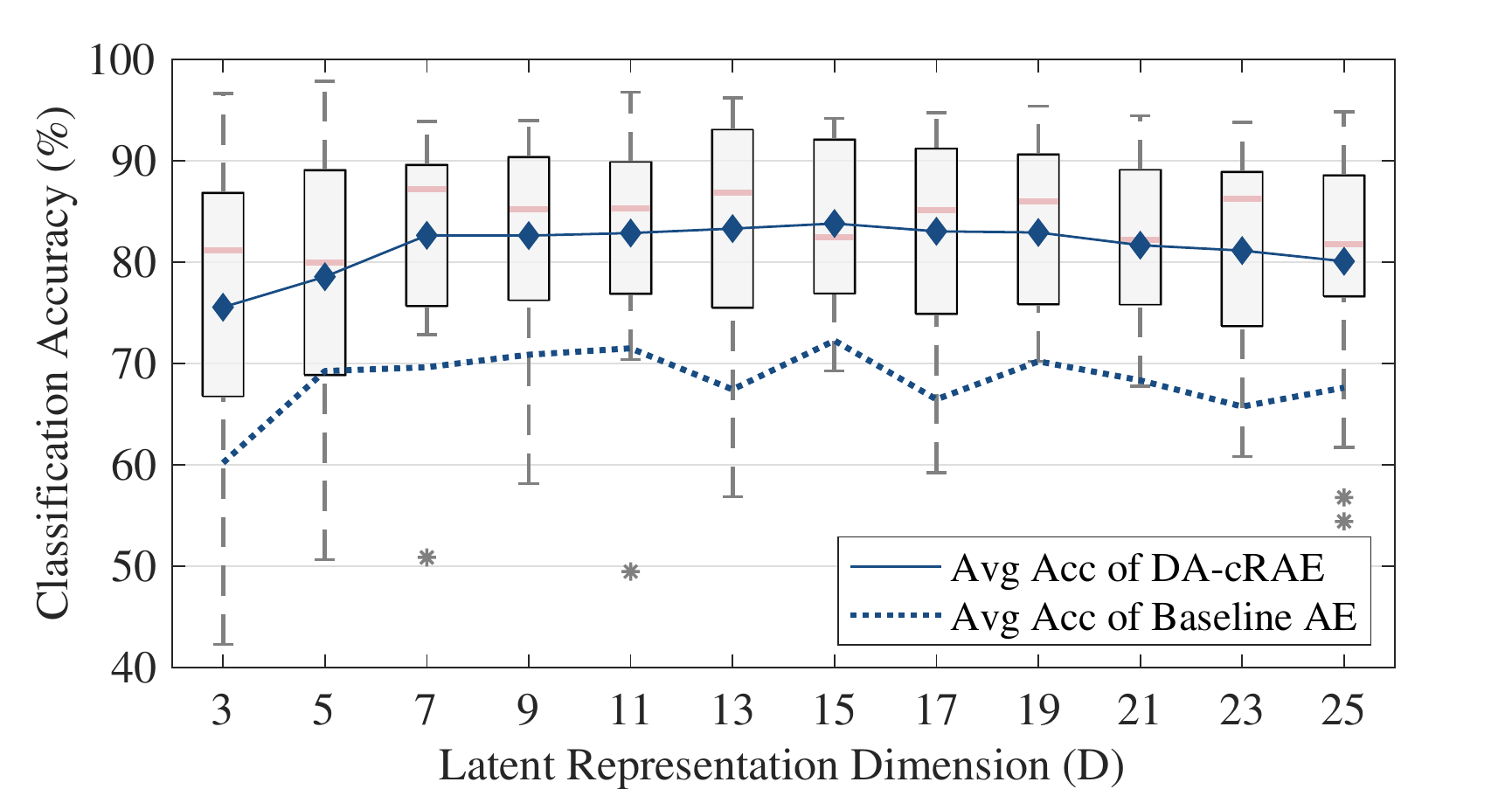}
  \caption{MLP classification accuracies of DA-cRAE model ($\lambda_A=0.5$, $\lambda_N=0.05$) for $20$ held-out subjects with different dimension D of representation $z$, compared with baseline AE.} %
\label{fig:acc-D}
\end{figure}

As shown in Fig.~\ref{results_box} and Table~\ref{params}, first we observe that simply feeding the decoder an extra conditional input $s$ could yield slightly better classification performance when comparing cAE with AE. 
Furthermore, we notice accuracy improvements from A-cAE and D-cAE to cAE, demonstrating that more cross-subject features observed in the hard-split representation $z_a$ lead to better identification of $y$. 
In addition, DA-cAE realizes further accuracy improvements with both adversary and nuisance networks compared to individual regularization approaches A-cAE and D-cAE. 
Under the disentangled adversarial transfer learning framework, our feature extractor results in lower variation of performances across all task classifiers and all subjects universally. 
More importantly, the soft-split RAE structures of A-cRAE, D-cRAE and DA-cRAE bring even more accuracy gain compared to the hard-split cases of A-cAE, D-cAE and DA-cAE. 
For the hard-split case, determining the split ratio of dimensions between subject-related and task-specified features is difficult since the representation nature is still unkown. 
However, the rateless property enables the encoder-decoder pair to seamlessly adjust dimensionalities of subject-related and task-specified features, and employs a smooth transition between the two stochastic counterparts by a probabilistic bottleneck representation, even though the underlying nature of the bottleneck is still vague. 
In general, the disentangled adversarial models of DA-cRAE with both adversary and nuisance networks attached to conditional decoder lead to significant improvements in average accuracy up to $11.6\%$ (e.g., the MLP classifier in Table~\ref{params}) with respect to the non-adversarial baseline AE. 
Furthermore, as observed in Fig.~\ref{results_box}, the cross-validation accuracies of the worst cases are also significantly improved, indicating that the proposed transfer learning architecture presents higher stability to a wider range of unknown individuals through reorganizing the subject- and task-relevant representations from the end of feature extractor.

\begin{figure}[t]
  \centering
  \includegraphics[clip,trim=0.3cm 0 0.8cm 0.1cm, width=\linewidth]{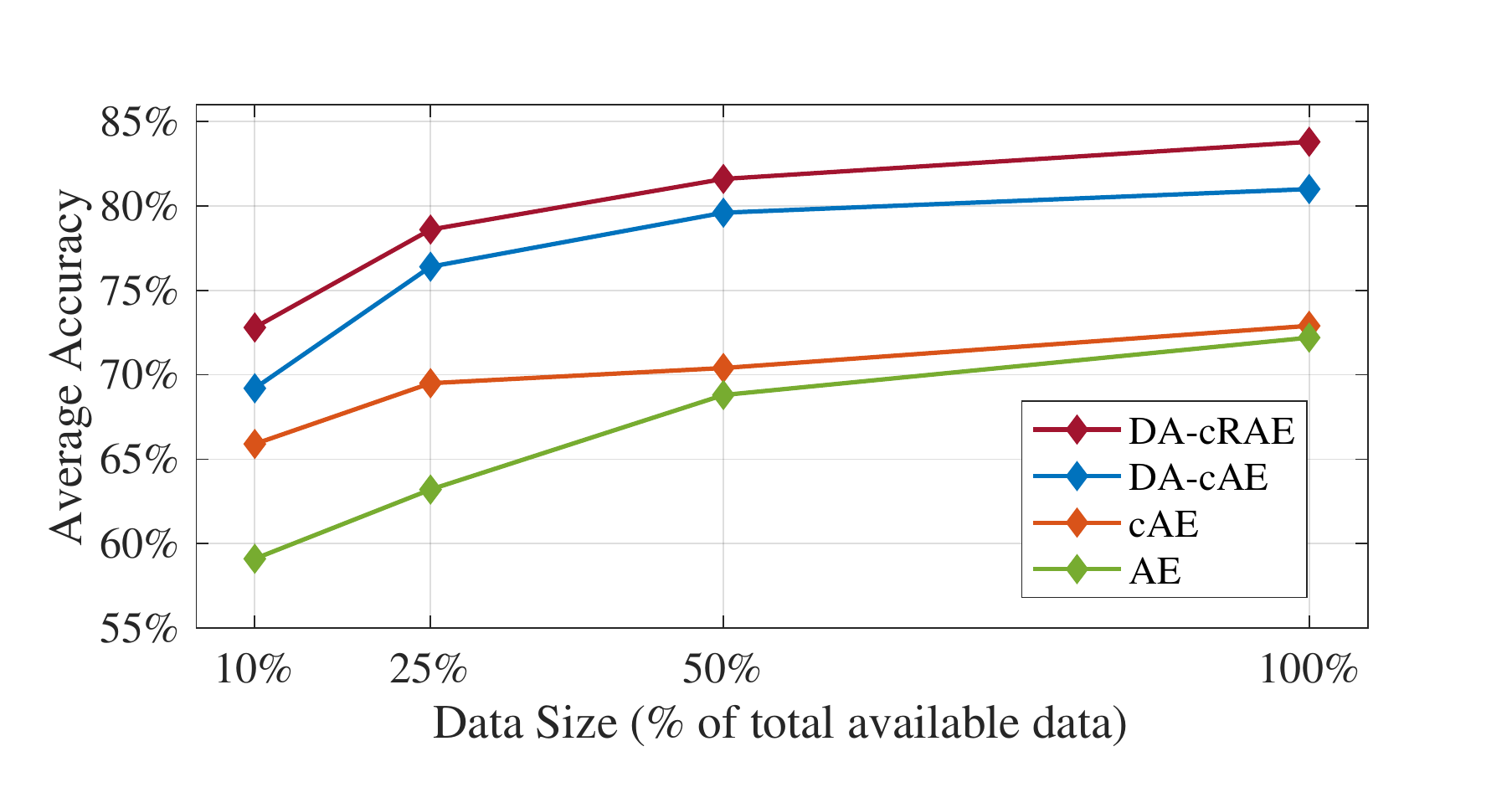}
  \caption{MLP classification accuracies of optimized parameter choices in Table~\ref{params} with different training dataset sizes.} 
\label{fig:datasize}
\end{figure}

\subsubsection{Impact of Disentangled Adversarial Parameters}
We take the MLP classifier as an example to particularly illustrate the impact of disentangled adversarial RAE. 
As presented in Table~\ref{results}, the baseline models of AE and cAE were first assessed with $\lambda_A=\lambda_N=0$ while training the MLP discriminative classifier.
Then the D-cRAE was evaluated with $\lambda_N \in \{ 0, 0.005, 0.01, 0.05, 0.2, 0.5 \}$ and $\lambda_A=0$. 
Finally, we froze $\lambda_N=0.05$ to observe the representation learning capability of the complete soft-disentangled adversarial transfer learning model DA-cRAE with different choices of $\lambda_A \in \{0, 0.01, 0.05, 0.1, 0.2, 0.5 \}$. 
For each parameter selection, the average accuracy of the MLP task classifier for identifying $4$ stress levels is shown in Table~\ref{results}, along with the discriminator accuracies of the adversary and nuisance blocks for decoding $20$-class ID.
With an increasing accuracy of MLP task classifier, stress levels are better discriminated; with a growing accuracy of nuisance network, more person-discriminative features are preserved in the nuisance counterpart; and with a decreasing accuracy of adversary network, more task-specific information are inherent in the adversary counterpart. 
We observe that the nuisance network produces higher accuracy with increasing $\lambda_N$, where $\lambda_N=0.05$ particularly results in the better performance on task classification. 
Furthermore, with fixed $\lambda_N=0.05$, growing $\lambda_A$ leads to lower accuracy of adversary network, and thus imposes less extraction of subject features but more task-related information on the adversary counterpart.

\subsubsection{Impact of Feature Dimension}
Other than the adversarial parameters $\lambda_A$ and $\lambda_N$, we further inspect the impact of different feature dimensions $D$ on the performance of the proposed DA-cRAE model. 
We trained MLP classifiers with the DA-cRAE feature extractor and its optimized parameters as given in Table~\ref{params} ($\lambda_A=0.5$ and $\lambda_N=0.05$), using various feature dimensions $D \in \{3,5,\cdots,25\}$. 
Corresponding cross-validation accuracies for $20$ held-out subjects are shown as a function of $D$ in Fig.~\ref{fig:acc-D}, where the average accuracy for each $D$ is also marked. 
The same assessments on $D$ were also applied to baseline AE feature extractor, and we present its curve of average accuracies in Fig.~\ref{fig:acc-D} as a reference to compare with DA-cRAE. 
It is verified that the proposed DA-cRAE consistently outperforms the baseline AE and $D=15$ latent dimensionality was sufficient for the problem.
We observe that after a specific value of dimension $D$, the performance of DA-cRAE remains relatively stable with varying $D$ value compared to AE. 
On one hand, when the feature dimension is large enough to carry necessary information for the classification task, higher $D$ value might not be able to bring more benefits when extracting features; on the other hand, the rateless property of DA-cRAE 
resolves the entanglement between task-related and subject-discriminative information and exploits the latent features in a more efficient manner, thus leading to a stronger robustness on the variance of latent representation dimensionality.

\subsubsection{Impact of Data Size}
In order to evaluate the robustness of our transfer learning method on data with smaller sizes, we investigated the performance of the proposed model when we reduced the available training data size from $100$\% to $50$\%, $25$\%, or $10$\%.
Corresponding classification accuracies as a function of training data size are shown in Fig.~\ref{fig:datasize}.
Here we consider the MLP classifier as the same example of Table~\ref{params}, to make comparisons among DA-cRAE ($\lambda_A=0.5$ and $\lambda_N=0.05$), DA-cAE ($\lambda_A=0.01$ and $\lambda_N=0.005$), cAE and AE ($\lambda_A=\lambda_N=0$). 
From Fig.~\ref{fig:datasize} we observe that DA-cRAE still performs best regardless of the amount of available training data. 
Even with $10$\% data only, there is no significant drawback of DA-cRAE and DA-cAE compared to non-adversarial methods, showing the transfer learning ability of our method to the size deficiency of physiological data. 
Note that with more available training data, even better performance is expected to be implemented by the model.

\begin{figure*}[t]
\centering
\subfloat[DA-cRAE convergence curves with 100\% data size.]{\includegraphics[clip, trim=0.2cm 0.3cm 0 1cm, width=0.49\textwidth]{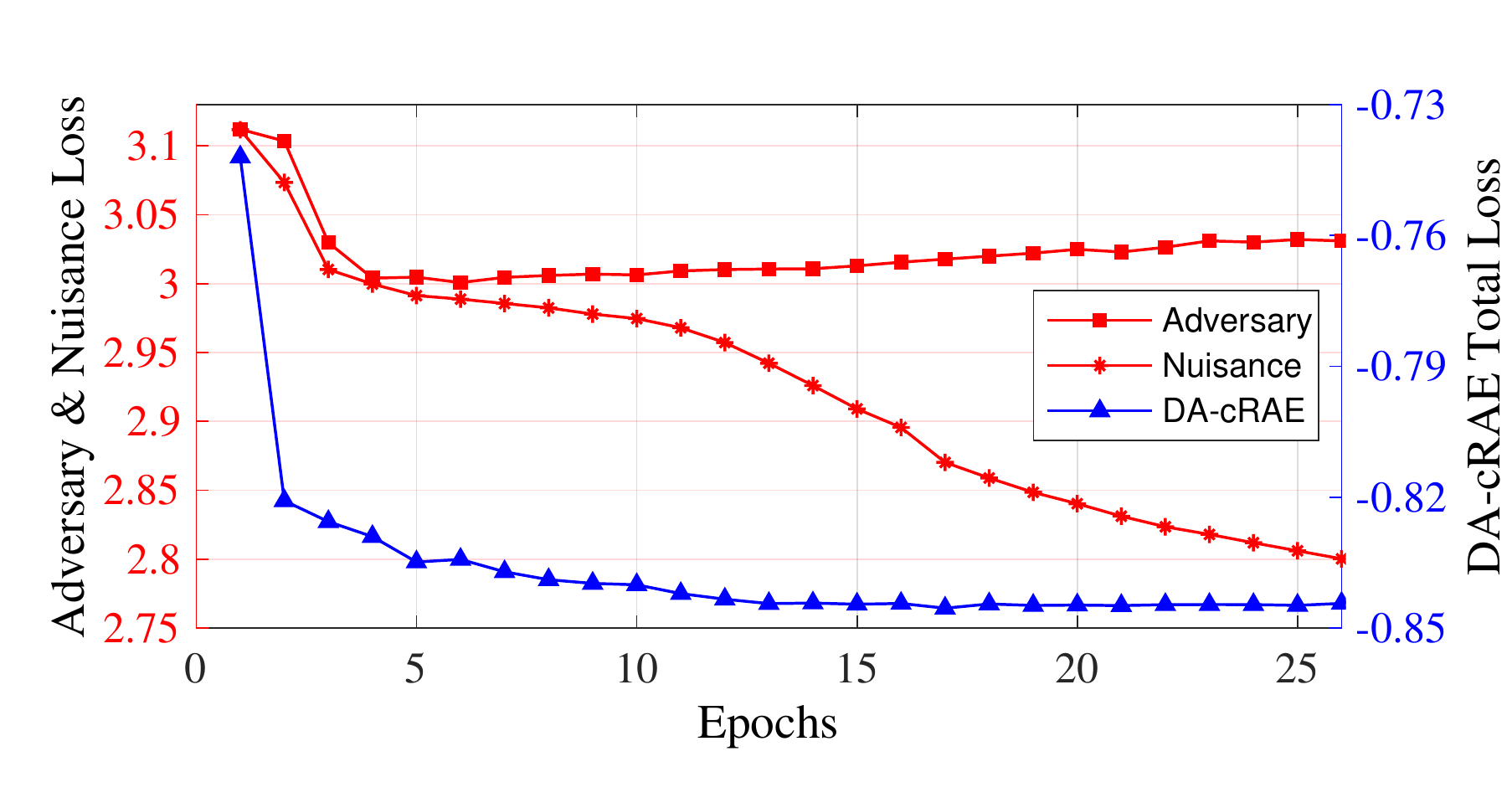}}
\subfloat[DA-cRAE convergence curves with 25\% data size.]{\includegraphics[clip, trim=0.2cm 0.3cm 0 1cm, width=0.49\textwidth]{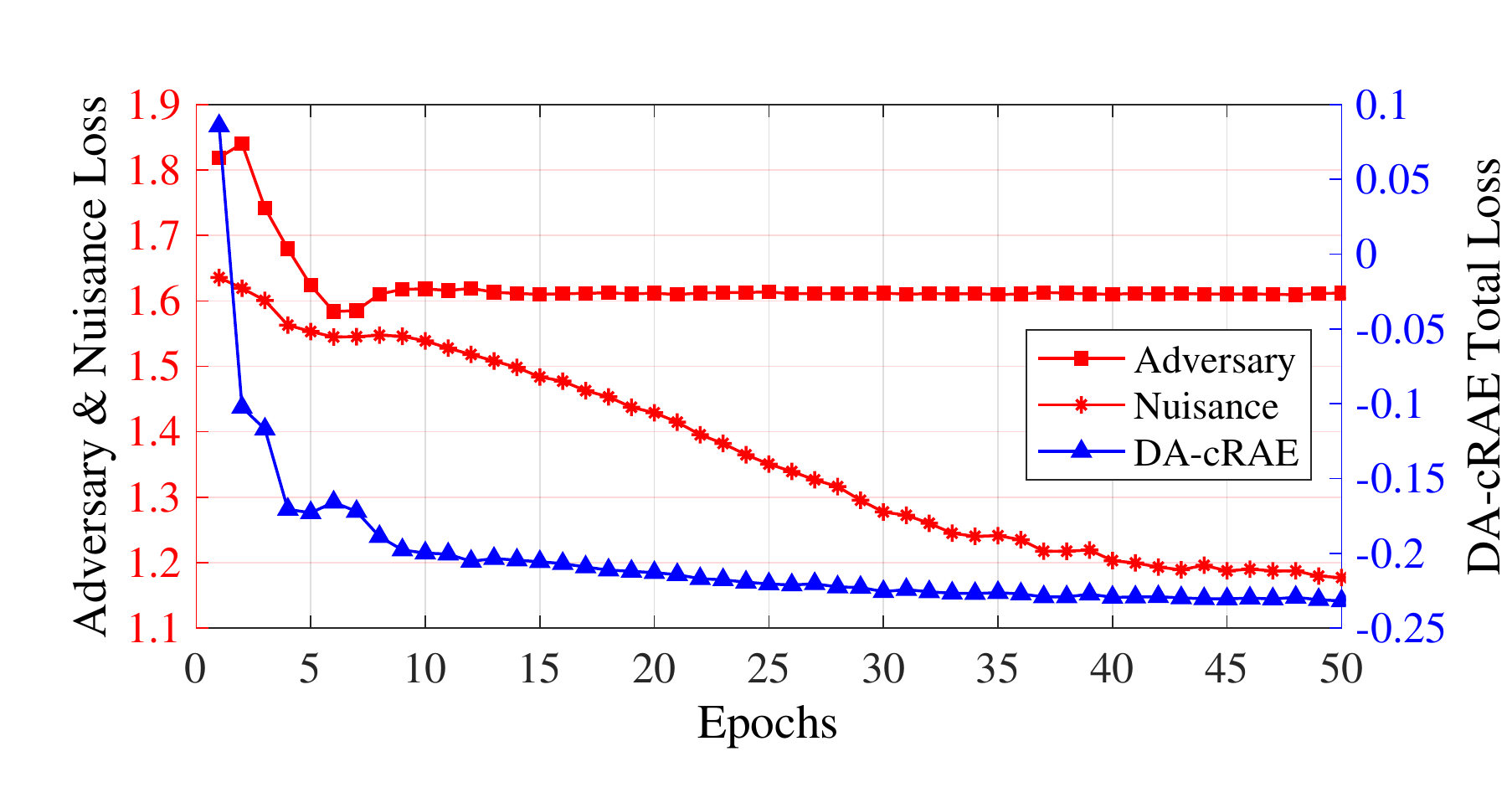}}
\caption{Convergence of DA-cRAE ($\lambda_A=0.5$ and $\lambda_N=0.05$) with different training data sizes.}
\label{fig:Convergence}
\end{figure*}

\subsubsection{Convergence Analysis}
In addition, training convergence curves for a specific DA-cRAE ($\lambda_A=0.5$ and $\lambda_N=0.05$) case with different training data sizes are presented in Fig.~\ref{fig:Convergence}. 
When using the full $100$\% set of available training data, i.e., in Fig.~\ref{fig:Convergence}(a), the total training loss value of DA-cRAE converges within $15$ epochs, 
while the nuisance loss decreases steadily with more training iterations and the loss value of the adversary unit keeps steady due to its antagonistic relationship with DA-cRAE, where the adversary unit continues to conceal subject-specific representations without undermining the discriminative performance of the entire network.
With less data, as illustrated in Fig.~\ref{fig:Convergence}(b), convergences are achieved after more training epochs, while the convergences of the DA-cRAE loss, adversary loss and nuisance loss are observed in a similar pattern with the full $100$\% data case, indicating the capability of the proposed model to learn universal features from data with even smaller sizes.
Overall, we observe that with both adversary and nuisance networks attached to the encoder, the classifier improves the accuracy substantially and shows more stable performance across different left-out subjects.

\section{Conclusion}

A transfer learning framework was proposed based on a soft-disentangled adversarial model utilizing the concept of RAE to extract universal and nuisance-robust physiological features. 
In order to implement the rateless property and manipulate the trade-off between subject-specific features and task-relevant information, additional blocks of adversary and nuisance networks were complementarily attached and jointly trained with different dropout strategies, and therefore the transfer learning framework is capable of handling a wider range of tasks and users. 
Cross-subject transfer evaluations were performed with a physiological biosignal dataset for monitoring human stress levels.
Significant benefits of the proposed framework were shown by improved worst-case accuracy and average classification accuracy, demonstrating the robustness to unknown users. 
The adaptability of the feature extractor over several task-discriminative linear and non-linear classifiers was also shown, and the transfer-learning ability of our method to data size deficiency was analysed.
Note that our methodology is applicable to various different systems requiring nuisance-robust analysis beyond HCI.


\bibliographystyle{IEEEtran}
\bibliography{refs}

\end{document}